\documentclass[aps,prb,showpacs
,twocolumn%
]{revtex4}
\usepackage[dvips]{graphicx}
\usepackage[hypertex]{hyperref}
\def\fr{f}
\def\fx{\tau}
\def\II{{\cal I}}
\makeatletter
\def\urlprefix#1#2{\hskip0pt plus0.01fil\discretionary{}{}{}%
  \hbox{\hyper@linkurl{online}{#2}}}
\begin{document}
\advance\textfloatsep by -0.3in
\advance\abovecaptionskip by -0.1in
\def\topfraction{1.00}
\def\textfraction{0.0}

\title{Driven classical diffusion with strong correlated disorder}
\date\today
 \author{Leonid P. Pryadko and Jing-Xian Lin}
\affiliation{Department of Physics, University of California,
  Riverside, California 92521}
\begin{abstract}
  We analyze one-dimensional motion of an overdamped classical
  particle in the presence of external disorder potential and an
  arbitrary driving force $F$.  In thermodynamical limit the effective
  force-dependent mobility $\mu(F)$ is self-averaging, although the
  required system size may be exponentially large for strong disorder.
  We calculate the mobility $\mu(F)$ exactly, generalizing the known
  results in linear response (weak driving force) and the perturbation
  theory in powers of the disorder amplitude.  For a strong disorder
  potential with power-law correlations we identify a non-linear
  regime with a prominent power-law dependence of the logarithm of
  $\mu(F)$ on the driving force.
\end{abstract}
\maketitle

\begin{figure}[ht]
  \begin{center}
    \includegraphics[width=0.95\columnwidth]{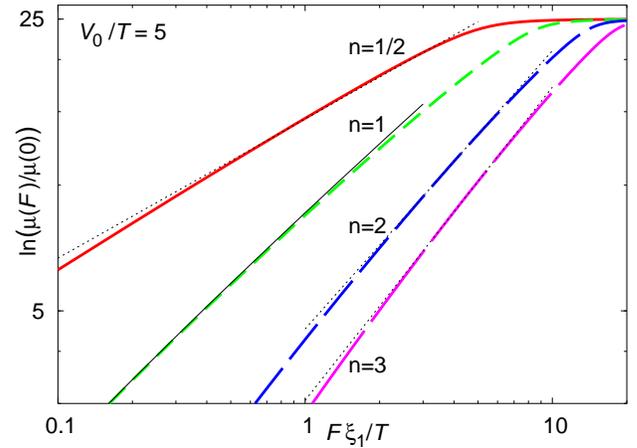}
    \caption{Logarithm of the effective mobility renormalization
      $\ln\biglb(\mu(F)/\mu(0)\bigrb)$
      [Eqs.~(\ref{eq:eta-average-one}),
        (\ref{eq:eta-average-small-drive})] with the correlation
      function $g(x)=(1+x^2/\xi_1^2)^{-n/2}$ for strong disorder
      ($V_0/T=5$) and large driving forces $F$.  Dotted lines guide
      the eye with the slope of the intermediate asymptote
      (\ref{eq:eta-strong-power-disorder-strong-drive-asymptotics}).
      Thin solid line is the analytic
      result~(\ref{eq:eta-strong-coulomb-disorder}) for $n=1$.}
    \label{fig:long-large}
  \end{center}
\end{figure}

\begin{figure}[hb]
  \begin{center}
    \includegraphics[width=\columnwidth]{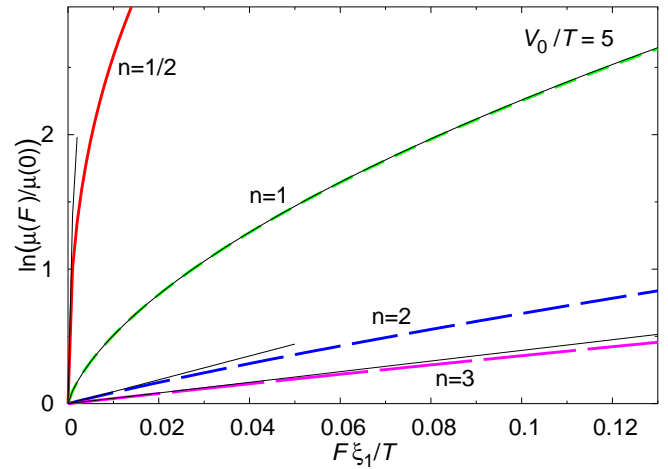}
    \caption{As in Fig.~\ref{fig:long-large} but for  small driving
      forces $F$.  Thin solid lines indicate the non-linear correction
      calculated analytically: linear in $F$ for $n>1$, proportional
      to $F^n$ for $n<1$, and
      Eq.~(\ref{eq:eta-strong-coulomb-disorder}) for $n=1$.}
    \label{fig:long-small}
  \end{center}
\end{figure}

We consider stationary diffusion in one dimension (1D), in the
presence of a random disorder potential and a constant driving field.
We show that the effective field-dependent mobility $\mu(F)$ is
self-averaging, and calculate its dependence on the field $F$ for
various forms of correlated disorder.  In particular, for a stationary
diffusion in a strong disorder potential with power-law correlations
at large distances [Eqs.~(\ref{eq:disorder-correlator}),
  (\ref{eq:def-power-law})], there is a wide intermediate range of
values of the driving force $F$ where the logarithm of the effective
mobility scales as a non-trivial power of $F$
(Fig.~\ref{fig:long-large}).  With Coulomb-like ($n=1$) or
longer-range correlations, the effective mobility is a singular
function of $F$ already at $F=0$; the applicability region of the
linear transport is essentially absent (Fig.~\ref{fig:long-small}).
These properties remain in the presence of weak interaction between
the particles, introduced here at the level of the self-consistent
Poisson equation which describes a Debye-like screening in the
presence of strong disorder.

Our results apply to a number of systems where classical
diffusion\cite{kramers-1940,risken-fpe,kramers-review-1990,%
  bouchaud-georges-1990} is the main mechanism of transport.  Other
applications include carrier diffusion in semiconductor
nanostructures, nano-scale thermoelectrics
\cite{Hicks-Dresselhaus-1993:zt-1D,lin-dresselhaus-prb-2003:te-nanowires},
transport in DNA\cite{hennig-dna} and other biological systems.  In
particular, the mean-field interaction model is directly applicable to
ionic transport through cell membranes\cite{hille-ionic-channels},
where one expects a number of parallel identical channels.

{\bf Single-particle diffusion} in 1D is described by the
Smoluchovsky equation
\begin{equation}
  \label{eq:langevin}
  \eta_0\dot x+\partial_xU(x)=f(t),
\end{equation}
where $U(x)$ is the external potential and $f(t)$ is the thermal force
with the correlator $\overline{ f(t)f(t')}=2 T\eta_0\delta(t-t')$.
The usual assumption is that the bare viscous friction coefficient
$\eta_0$ is determined by fast scattering events off phonons,
short-range disorder, etc.  The potential $U(x)$ in
Eq.~(\ref{eq:langevin}) is thus the part of the overall potential
remaining after averaging over some distance scale; its precise value
depends on the specific physical system.  If we assume the carriers
have charge $e$, we can also define the bare mobility in the absence
of disorder, $U(x)=-eEx$:
\begin{equation}
  \label{eq:bare-mobility}
  \mu_0\equiv \overline{\dot x}/E=e/\eta_0.
\end{equation}

The equation~(\ref{eq:langevin}) can be also rewritten as the
transport equation for the average particle density $n\equiv n(x,t)$
and the particle current $j\equiv j(x,t)$,
\begin{equation}
\partial_t n+\partial_x
j=0,\quad j=-D_0 \partial_x n-
\eta_0^{-1}  n\,\partial_x U(x);\label{eq:diffusion}
\end{equation}
the diffusion constant $D_0$ is related to the viscous
friction coefficient $\eta_0$ by the Einstein relation,
$D_0=T/\eta_0$.

The usual transport problem corresponds to stationary diffusion in the
presence of a random Gaussian potential $V(x)$ and a constant driving
force $F$, with the total potential energy $U(x)=V(x)-Fx$.  In the
case of a periodic potential $V(x)=V(x+a)$, the stationary
solution\cite{kramers-1940,risken-fpe,kramers-review-1990} corresponds to a
constant average current $j$ with periodic boundary conditions,
$n(a)=n(0)$.  Then, by normalizing the density profile $n(x)$ over the
period, we can use the current to define the average drift velocity,
$\bar v=j/\bar n=j a$, as well as the effective viscous friction
coefficient $\eta\equiv F/\bar v= {F/ j a}$,
\begin{equation}
  \label{eq:eta}
  \eta 
  ={\eta_0 F\over a\, T}\!
  \int_0^a \!\!\! dx \int_{x}^{x+a}\!\!\!\!\! dx'\,
  {e^{[V(x')-V(x)+F(x-x')]/T}\over
  1-e^{-Fa/T}}.
\end{equation}
Both integrations extend over the entire period, and, as it often
happens in classical transport phenomena, for a sufficiently large $a$
the effect of disorder becomes self-averaging [the required size can
  be large, see Eq.~(\ref{eq:relative-eta2-strong-disorder})
  below].  In such cases we can replace Eq.~(\ref{eq:eta}) by its
average over disorder.  We assume that the disorder distribution is
Gaussian with the correlations
\begin{equation}
  \label{eq:disorder-correlator}
  \langle V(x)\rangle=0,\quad
  \langle V(x) V(x')\rangle=V_0^2 g(x-x'),
\end{equation}
where the local r.m.s.\ value $V_0$ is taken as the measure of the
disorder strength and the correlation function $g(x)$ is defined so
that $g(0)\equiv 1$.  Then, in the thermodynamical limit $a\to\infty$,
the effective mobility $\mu(F)\equiv e/\langle\eta\rangle$, and
\begin{equation}
  \label{eq:eta-average-one}
{\mu_0\over \mu(F)}
  =e^{V_0^2/T^2}{F\over T}\!\int_0^\infty\!\!\!\!
  dx\,e^{-Fx/T}\,e^{-g(x)V_0^2/T^2}\!,\;\, F>0.\!
\end{equation}
In the small-$F$ limit (but at the same time \hbox{$aF\gg T$}), this
gives the usual linear response
result\cite{festa-dagliano-1978,Alexander-Orbach-1981,%
  golden-1985,schneider-1988},
\begin{equation}
  \langle\eta\rangle/\eta_0\stackrel{F\to0}= \mu_0/\mu(0)
  =\exp({V_0^2/T^2}),
  \label{eq:eta-average-small-drive}
\end{equation}
which can be understood as the average of the activation exponent of the
difference between the highest maximum and the lowest minimum of the
potential.  Generically, these would be in different parts of the
sample and so the friction renormalization factors onto a product of
the two averages $\langle e^{V/T} \rangle \langle
e^{-V/T}\rangle=\exp(V_0^2/T^2)$, independent of the form of the
correlation function $g(x)$.

Similarly, the stationary limit of the dynamical perturbation
theory\cite{larkin-ovchinnikov-1974,vinokur-vortex-liquid-1990}
is restored by expanding
Eq.~(\ref{eq:eta-average-one}) in powers of $V_0/T$ and integrating
the result by parts,
\begin{equation}
  \label{eq:eta-disorder-expansion}
{\mu_0\over \mu(F)}=1-{V_0^2\over T^2}\int_0^\infty\!\!\!
  dx\,e^{-Fx/T}\,g'(x)+{\cal O}(V_0^4/T^4).
\end{equation}
The disorder correlation function $g(x)$ is expected to decrease with
$x$, remaining substantially different from zero over the distance of
the order of the appropriate correlation length $\xi$.  It is clear
from the weak-disorder expression~(\ref{eq:eta-disorder-expansion})
that for such finite-range disorder there is a distinct crossover
force $F_\xi\sim T/\xi$: while $F\alt F_\xi$ have relatively little
effect on the mobility, larger values of $F$
begin to suppress the effect of disorder as large-scale potential
valleys and hills gradually disappear. 

To analyze an analogous effect for {\em strong\/} finite-range
disorder, note that for large $V_0/T$, $\Delta_x\equiv e^{-g(x)
  V_0^2/T^2}$ is exponentially small for $x\alt\xi$.  This effectively
limits the integration in Eq.~(\ref{eq:eta-average-one}) to the region
$x\agt \xi$, so that
\begin{equation}
  \label{eq:eta-strong-short-disorder}
  \ln\biglb(\mu(F)/\mu(0)\bigrb)\sim {F\tilde\xi/ T},
\end{equation}
where $\tilde\xi\approx \xi$ up to a logarithmic correction.  Such a
dependence on the applied field is analogous to the logarithmic
susceptibility\cite{smelyanskiy-dykman-1997} typical for systems with
activated transport.  Here it can be understood as the diffusion
limited by far-spaced maxima of the potential, with the particles
concentrated in the intermediate low minima; the applied driving force
$F$ effectively reduces the energy gap between the minima and the
maxima and therefore has an {\em exponential\/} effect on the
mobility.

The expression~(\ref{eq:eta-strong-short-disorder}) is valid
qualitatively as long as the effect of the disorder remains large,
$\mu_0/\mu(F)\gg1$.  The precise value of $\tilde\xi$ and the prefactor
depends on the details of the disorder correlation function.  For
example, with the exponential correlation function
$g(x)=\exp(-x/\xi)$, the integration (\ref{eq:eta-average-one}) can be
done exactly in terms of the incomplete gamma function; the asymptotic
form for $V_0^2/T^2\gg \max(1,\chi\equiv F\xi/T)$ is
\begin{equation}
  \label{eq:eta-strong-exp-disorder}
  {\mu(F)\over \mu(0)}
  =    {(V_0/T)^{2\chi}\over \Gamma(\chi+1)}\,\stackrel{\chi\agt1}\to\,
  (2\pi\chi)^{-1/2}\, e^{F\tilde\xi/T},
\end{equation}
where $\tilde\xi=\xi \bigl[1+\ln\biglb(V_0^2/(T F\xi)\bigrb)\bigr]$.

The disorder-induced transport non-linearity becomes even more
pronounced for long-range potentials, e.g., those with power-law
correlations at large distances.  With long-range correlations the
far-spaced maxima and minima of the potential are not entirely
independent and, therefore, even a weak driving force may have a
noticeable effect.  Specifically, consider a correlation function with
the asymptotic form
\begin{equation}
  g(x)= (\xi_1/x)^n,\quad
  x>x_{\rm min}\gg \xi,\xi_1.\label{eq:def-power-law}
\end{equation}
With strong
enough disorder [$g(x_{\rm min}) V_0^2/T^2\agt1$] the
integral~(\ref{eq:eta-average-one}) will be determined by large $x$,
in which case the expression can be rewritten approximately as
\begin{equation}
  \label{eq:eta-strong-power-disorder}
{\mu(0)\over \mu(F)}\approx
 \II_n(\alpha),\;\,\II_n(\alpha)\equiv \int_0^\infty\!\!\!\!
  dx\, e^{-x-\alpha x^{-n}},
\end{equation}
where $\alpha\equiv \bigl({F\xi_1/ T} \bigr)^n {V_0^2/T^2}$ can be
small or large within the strong-disorder domain where
Eq.~(\ref{eq:eta-strong-power-disorder}) is applicable.  For
sufficiently large $\alpha$, the integration can be done using the
Gaussian approximation around the maximum at $x_0=(n\alpha)
^{1/(n+1)}$,
\begin{equation}
  \label{eq:eta-strong-power-disorder-strong-drive}
  \II_n(\alpha)  =
  \Bigl({2\pi x_0 \over
    n+1}\Bigr)^{1/2} e^{-  x_0
    (1+1/n)},\;\,\alpha\agt1.
\end{equation}
As a result, logarithm of $\mu(F)$ is proportional to a power,
\begin{equation}
  \label{eq:eta-strong-power-disorder-strong-drive-asymptotics}
    \ln\biglb(\mu(F)/\mu(0)\bigrb)\sim
    \bigl({F\tilde\xi_1/ T}\bigr)^{n/(n+1)},
\end{equation}
with a {\em large\/} temperature-dependent length parameter
$\tilde\xi_1=C_n\xi_1\,(V_0/T)^{2/n}$, $C_n\equiv (n+1)^{1+1/n}/n$
[cf.\ Eq.~(\ref{eq:eta-strong-short-disorder})].  For small
$\alpha\ll1$, the integration~(\ref{eq:eta-strong-power-disorder}) can
be done perturbatively in powers of $\alpha\propto F^n V_0^2$ for
$n<1$ or, using the identity $\II_n(\alpha)=\II_{1/n}(\alpha^{1/n})$, in
powers of $\alpha^{1/n}\propto FV_0^{2/n}$ for $n>1$.  For Coulomb
disorder, $n=1$, the result is expressed in terms of the Macdonald
function,
\begin{equation}
  \label{eq:eta-strong-coulomb-disorder}
  \mu(0)/\mu(F)\bigr|_{n=1}\approx
  \II_1(\alpha)=2\alpha^{1/2} K_1(2\alpha^{1/2}).
\end{equation}
For very small $\alpha\ll1$, $\II_1(\alpha)\approx 1-\alpha
\ln\bigl({e^{1-2\gamma}/\alpha}\bigr)$, where $\gamma\approx 0.577$ is
the Euler's constant; the correction is linear in $F$ up to a
logarithm.  Clearly, for strong Coulomb or longer-range disorder,
$n\le 1$, the mobility $\mu(F)$ is a singular
function of the driving force at $F=0$; the linear-transport regime is
essentially absent.  These asymptotics are illustrated in
Figs.~\ref{fig:long-large}, \ref{fig:long-small} for a model form of
the disorder correlation function $g(x)=(1+x^2/\xi_1^2)^{-n/2}$.

{\bf Self-averaging}.  Our conclusions on the scaling of mobility in
strongly-disordered diffusive 1D systems are based on the average,
Eq.~(\ref{eq:eta-average-one}).  To analyze the sample-to-sample
fluctuations, consider the irreducible average
$\langle\!\langle\eta^2\rangle\!\rangle\equiv
\langle\eta^2\rangle-\langle\eta\rangle^2$ of the effective friction
$\eta$ [Eq.~(\ref{eq:eta})],
\begin{eqnarray}
  \nonumber
  \lefteqn{{\langle\!\langle\eta^2\rangle\!\rangle\over\eta_0^2}=
    e^{2V_0^2/T^2}{F^2\over a\,T^2}\int_0^a\!\!\!\!
    dx\,e^{-Fx/T}\Delta_x\int_0^a\!\!\!\! dy\,e^{-Fy/T}\Delta_y}&&
  \hskip0.85\columnwidth\\
  &\times&\int_0^{a}\!\!\!\!
  dz \,  \Bigl(
  \Delta_{z+{x+y\over2}}\Delta_{z-{x+y\over2}}
  \Delta^{-1}_{z+{x-y\over 2}}\Delta^{-1}_{z-{x-y\over2}}-1
  \Bigr),
  \label{eq:eta-fluct-average-one}
\end{eqnarray}
where the correlator $\Delta_z\equiv e^{-g(z)V_0^2/T^2}$ is periodic
under $z\to z+a$.  Note that both $\Delta_z$ and $\Delta_z^{-1}$ enter
Eq.~(\ref{eq:eta-fluct-average-one}).  Therefore, unlike in the
average~(\ref{eq:eta-average-one}), both short- and long-distance
disorder correlations affect the variance of $\eta$.

For weak disorder, the expansion of
Eq.~(\ref{eq:eta-fluct-average-one}) in powers of $V_0/T\ll1$ begins
with the quartic term,
\begin{eqnarray}
  \lefteqn{{\langle\!\langle\eta^2\rangle\!\rangle\over\eta_0^2}=
    {V_0^4\over T^4}\!{F^2\over a T^2}\,\int_0^a\!\!\!\! dx\, e^{-Fx/T}
    \!\!\int_0^a\!\!\!\! dy\, e^{-Fy/T}\!\!\int_0^a\!\!\!\! dz
    }& &\hskip3.0in \nonumber\\
  &&\hskip0.2in \times\, g_{z}\,
  \bigl(2 g_{z}+g_{z+x+y}+g_{z+x-y}-2 g_{z+x}-2 g_{z+y}\bigr)\nonumber \\
  &=&{2V_0^4\over a T^4}\int_0^a \!\!\!\! dz\,g_z \!\!\int_0^a \!\!\!\! du
    \,e^{-Fu /T}\,(u g''_{z+u}-g'_{z+u}).
  \label{eq:eta-fluct-weak-disorder}
\end{eqnarray}
The integration is simplified in the limits of weak and large $F$; the
combined result is
\begin{equation}
  \label{eq:eta-fluct-weak-disorder-result}
  {\langle\!\langle\eta^2\rangle\!\rangle\over\eta_0^2}
    ={V_0^4\over 2 a  T^4}\min(4\xi_2,T/F),\;\,
    \xi_2\equiv
    \int_0^\infty\!\!\!\! dx\,g^2(x).
\end{equation}
Here, $\xi_2$ is yet another correlation length, finite for
short-range disorder and for long-range disorder with $n>1/2$.
Clearly, for weak disorder the variation of $\eta$ is small and it
is further reduced with increasing system size $a$.

The situation is different for strong disorder, $V_0\gg T$, which
causes an exponential renormalization of $\eta$; large fluctuations
are also expected.  In this case $\Delta^{-1}_x$ has a prominent
maximum at the origin.  Consequently, the
integral~(\ref{eq:eta-fluct-average-one}) gets an exponentially large
contribution from a vicinity of the point $z=0$, $x=y$.  Using the
steepest descent method, we obtain
\begin{equation}
  \label{eq:eta-fluct-average-two}
  {\langle\!\langle\eta^2\rangle\!\rangle\over \eta_0^2}
  =  {2\pi F^2 e^{4V_0^2/T^2}\over a V_0^2}\!\int_0^\infty\!\!\!\!du\, {
    e^{-2F u/T}e^{-4 g(u)V_0^2/T^2}\over  g''(u)-g''(0)}.\!
\end{equation}
For not exceedingly large $F$ the result is determined by values of
$u$ away from the origin.  Then, the denominator can be replaced by a
constant\footnote{Here $\xi_0$ is a correlation length describing the
  short-distance properties of the disorder potential; it may be quite
  different from the length $\xi$ relevant at large distances.}
$-g''(0)\equiv 2/\xi_0^2$, and the integral acquires precisely the
form of Eq.~(\ref{eq:eta-average-one}).  Generally,
\begin{equation}
{\delta\mu^2\over \mu^2}\approx
{\langle\!\langle \eta^2\rangle\!\rangle\over\langle\eta\rangle^2}
=C\,
{\pi F\, T\xi_0^2\over 2 a \,V_0^2}\,{e^{2V_0^2/T^2}};
\label{eq:relative-eta2-strong-disorder}
\end{equation}
while $F$ in the prefactor can be small, it is assumed to be large on
the scale of the system size, $Fa/T\gg1$.  The normalization in
Eq.~(\ref{eq:relative-eta2-strong-disorder}) is chosen so that for
short-range disorder $C\approx1$ [cf.\ 
  Eq.~(\ref{eq:eta-strong-short-disorder})].  For power-law
correlation tail, $ C={\II_n(2^{n+2}\alpha)/ [\II_n(\alpha)]^2}$, with
$\alpha\equiv (F\xi_1/T)^nV_0^2/T^2$ [cf.\ Eqs.~(\ref{eq:def-power-law}),
  (\ref{eq:eta-strong-power-disorder})].  Overall, we conclude that
the effective mobility is a self-averaging quantity in the
thermodynamical limit.  Of course, the required system size can be
large if the fluctuations are strong.

We verified these conclusions by simulating diffusion in a 1D
short-range random potential (not shown).  With periodic boundary
conditions the viscous friction~(\ref{eq:eta}) could be obtained by
averaging the time $t$ it takes a particle to travel over one period,
$\Delta x=a$.  As expected, with increasing $a$, the corresponding
disorder average $\langle t\rangle/a$ approached the inverse of the
average drift velocity.

{\bf Stationary diffusion with weak interaction}.  The considered
problem differs from the canonical Kramers
problem\cite{kramers-1940,risken-fpe,kramers-review-1990} of
over-the-barrier transport, where it is the dynamical equilibrium that
establishes the exponentially different particle numbers in the
``reservoirs'' on the two sides of the barrier.  Here, we consider a
situation corresponding to a typical resistivity measurement in a
macroscopic sample where the total number of particles does not change
with the applied field.  Then, the macroscopic current would be
determined solely by the average drift velocity.  It is important that
the quantity is self-averaging, as the explicit disorder averaging
would not be necessary for large enough samples.  Previously,
stationary driven diffusion was considered within the linear
response  with random disorder\cite{%
  festa-dagliano-1978,%
  Alexander-Orbach-1981,%
  golden-1985,%
  schneider-1988}, and also for arbitrary $F$ in {\em deterministic\/}
periodic
potentials\cite{risken-fpe,kramers-review-1990,%
  stepanov-sommer-1990,stepanov-1991}.

In finite-size systems such a situation arises naturally, e.g., when
diffusing particles are charged and the electroneutrality condition
needs to be satisfied.  The simplest case corresponds to the Debye
mean-field screening, where the potential in Eq.~(\ref{eq:diffusion})
is modified by the self-consistent potential, $U(x)\to
U(x)+e\varphi(x)$.  Specifically, we consider a 1D Poisson equation,
\begin{equation}
  \varphi''=-4\pi e (n-\bar n),\label{eq:poisson}
\end{equation}
as would be appropriate for diffusion in a 3D system with
1D-modulation (layered disorder), a parallel bunch of identical DNA
molecules, or electrostatically-coupled ionic cell channels.

In the static equilibrium, $F=0$, the coupled Eqs.~(\ref{eq:diffusion}),
(\ref{eq:poisson}) correspond to the non-linear screening problem;
with weak disorder, $V_0\ll T$, the Debye screening length is
$\kappa^{-1}$, $\kappa^2\equiv 4\pi \bar n e^2/T$.  The linearized
self-consistent screening problem can be also solved with the non-zero
driving force, $F>0$; the solution involves two screening parameters,
$s_\pm=(\kappa^2+\fr^2/4)^{1/2}\pm \fr/2$, where $\fr\equiv F/T\approx
j/(D_0\bar n)\equiv \lambda$.  Clearly, in the weak-interaction limit,
$\fr\gg \kappa$, the shorter screening length $s_+^{-1}\approx
\fr^{-1}=T/F$ is determined by the driving force, while the longer one
diverges, $s_-^{-1}\approx \fr/\kappa^2$.

With the driving force and a strong disorder, the problem is
forbiddingly complicated.  However, if the interaction is weak, the
additional potential would be small, and the screening equations can
be linearized.  To this end, it is convenient to eliminate the density
$n$ from Eqs.~(\ref{eq:diffusion}), (\ref{eq:poisson}) and write the
self-consistent equation for the scaled gradient of the screening
potential $\varepsilon\equiv e\varphi'(x)/T$,
$$
  \label{eq:self-consistent}
  \varepsilon''+\varepsilon' \Bigl({V'\over
    T}-\fr\Bigr)-\kappa^2\varepsilon=\fx -\varepsilon\varepsilon',\;
  \fx\equiv \kappa^2\Bigl({V'\over T}+\lambda-\fr\Bigr).
$$
This equation can be rewritten identically as
$$
e^{-V/T}\Bigl({d\over dx}-s_+\Bigr)e^{V/T}
\Bigl(\varepsilon'+s_-\varepsilon\Bigr)=\fx-\Bigl(\varepsilon'-s_-{V'\over
  T}\Bigr)\varepsilon.
$$
For weak interaction, the last term in the r.h.s.\ is quadratic in
small $\kappa^2$ and can be ignored.  The remaining equation should be
solved for $\varepsilon$ with zero boundary conditions at infinity.
The relation between the driving force $F\equiv \fr T$ and the
diffusion current $j\equiv\lambda\bar n D_0$ is established from the
self-consistency condition that the screening does not modify the net
driving field.  Equivalently, the disorder-averaged
$\langle\varepsilon\rangle=0$.  Approximating $s_+\approx\fr$, after
some algebra we again arrive at
Eq.~(\ref{eq:eta-average-one}).

{\bf Conclusions.}  We analyzed the stationary 1D problem of a driven
diffusion in the presence of a random disorder potential.  For large
systems and/or in the presence of an interaction fixing the number of
particles transport should be described in terms of the effective
mobility $\mu$.  Strong disorder significantly reduces the mobility
and leads to its non-trivial scaling as a function of the driving
field $F$ and the temperature.  With finite-range disorder, the
dependence $\mu(F)$ can be described in terms of the logarithmic
susceptibility, Eq.~(\ref{eq:eta-strong-short-disorder}), a generic
form for problems with activated transport.  For a strongly-driven
system with power-law disorder correlations the logarithm of the
mobility scales as a power of the driving force,
Eq.~(\ref{eq:eta-strong-power-disorder-strong-drive-asymptotics}).
The disorder effect is especially pronounced for Coulomb-like
correlations [see Eq.~(\ref{eq:eta-strong-coulomb-disorder})]: the
field-dependent correction to mobility is singular already at $F=0$.

The obtained explicit results are applicable for a number of systems
where diffusion is the main transport mechanism, and can be especially
useful for characterization of disorder distribution.  Both quantum
effects and particle-particle interaction would further modify the
functional form of mobility.  For the purposes of comparison with
experiment at low temperatures, it would be useful to obtain the
corresponding results beyond the usual asymptotic limits of zero
temperature and delta-correlated disorder.

{\bf Acknowledgments}.  The authors are grateful to Mark Dykman and
Chandra Varma for encouragement and illuminating discussions.

\bibliography{diffus,scvort}

\begin{thebibliography}{17}
\expandafter\ifx\csname natexlab\endcsname\relax\def\natexlab#1{#1}\fi
\expandafter\ifx\csname bibnamefont\endcsname\relax
  \def\bibnamefont#1{#1}\fi
\expandafter\ifx\csname bibfnamefont\endcsname\relax
  \def\bibfnamefont#1{#1}\fi
\expandafter\ifx\csname citenamefont\endcsname\relax
  \def\citenamefont#1{#1}\fi
\expandafter\ifx\csname url\endcsname\relax
  \def\url#1{\texttt{#1}}\fi
\expandafter\ifx\csname urlprefix\endcsname\relax\def\urlprefix{URL }\fi
\providecommand{\bibinfo}[2]{#2}
\providecommand{\eprint}[2][]{\url{#2}}

\bibitem[{\citenamefont{Kramers}(1940)}]{kramers-1940}
\bibinfo{author}{\bibfnamefont{H.~A.} \bibnamefont{Kramers}},
  \bibinfo{journal}{Physica} \textbf{\bibinfo{volume}{VII}},
  \bibinfo{pages}{284} (\bibinfo{year}{1940}).

\bibitem[{\citenamefont{Risken}(1984)}]{risken-fpe}
\bibinfo{author}{\bibfnamefont{H.}~\bibnamefont{Risken}},
  \emph{\bibinfo{title}{The Fokker-Planck Equation}}
  (\bibinfo{publisher}{Springer}, \bibinfo{address}{Berlin},
  \bibinfo{year}{1984}).

\bibitem[{\citenamefont{H\"anggi et~al.}(1990)\citenamefont{H\"anggi, Talkner,
  and Borkovec}}]{kramers-review-1990}
\bibinfo{author}{\bibfnamefont{P.}~\bibnamefont{H\"anggi}},
  \bibinfo{author}{\bibfnamefont{P.}~\bibnamefont{Talkner}}, \bibnamefont{and}
  \bibinfo{author}{\bibfnamefont{M.}~\bibnamefont{Borkovec}},
  \bibinfo{journal}{Rev. Mod. Phys.} \textbf{\bibinfo{volume}{62}},
  \bibinfo{pages}{251} (\bibinfo{year}{1990}),
  \urlprefix\url{http://link.aps.org/abstract/RMP/v62/p251}.

\bibitem[{\citenamefont{Bouchaud and Georges}(1990)}]{bouchaud-georges-1990}
\bibinfo{author}{\bibfnamefont{J.-P.} \bibnamefont{Bouchaud}} \bibnamefont{and}
  \bibinfo{author}{\bibfnamefont{A.}~\bibnamefont{Georges}},
  \bibinfo{journal}{Physics Reports} \textbf{\bibinfo{volume}{195}},
  \bibinfo{pages}{127} (\bibinfo{year}{1990}),
  \urlprefix\url{http://dx.doi.org/10.1016/0370-1573(90)90099-N}.

\bibitem[{\citenamefont{Hicks and
  Dresselhaus}(1993)}]{Hicks-Dresselhaus-1993:zt-1D}
\bibinfo{author}{\bibfnamefont{L.~D.} \bibnamefont{Hicks}} \bibnamefont{and}
  \bibinfo{author}{\bibfnamefont{M.~S.} \bibnamefont{Dresselhaus}},
  \bibinfo{journal}{Phys. Rev. B} \textbf{\bibinfo{volume}{47}},
  \bibinfo{pages}{16631} (\bibinfo{year}{1993}),
  \urlprefix\url{http://link.aps.org/abstract/PRB/v47/p16631}.

\bibitem[{\citenamefont{Lin and
  Dresselhaus}(2003)}]{lin-dresselhaus-prb-2003:te-nanowires}
\bibinfo{author}{\bibfnamefont{Y.-M.} \bibnamefont{Lin}} \bibnamefont{and}
  \bibinfo{author}{\bibfnamefont{M.~S.} \bibnamefont{Dresselhaus}},
  \bibinfo{journal}{Phys. Rev. B} \textbf{\bibinfo{volume}{68}},
  \bibinfo{pages}{075304} (\bibinfo{year}{2003}),
  \urlprefix\url{http://link.aps.org/abstract/PRB/v68/e075304}.

\bibitem[{\citenamefont{Hennig et~al.}(2003)\citenamefont{Hennig, Archilla, and
  Agarwal}}]{hennig-dna}
\bibinfo{author}{\bibfnamefont{D.}~\bibnamefont{Hennig}},
  \bibinfo{author}{\bibfnamefont{J.~F.~R.} \bibnamefont{Archilla}},
  \bibnamefont{and} \bibinfo{author}{\bibfnamefont{J.}~\bibnamefont{Agarwal}},
  \bibinfo{journal}{Physica D} \textbf{\bibinfo{volume}{180}},
  \bibinfo{pages}{256} (\bibinfo{year}{2003}).

\bibitem[{\citenamefont{Hille}(1992)}]{hille-ionic-channels}
\bibinfo{author}{\bibfnamefont{B.}~\bibnamefont{Hille}},
  \emph{\bibinfo{title}{Ionic Channels of Excitable Membranes}}
  (\bibinfo{publisher}{Sinauer}, \bibinfo{address}{Sunderland, MA},
  \bibinfo{year}{1992}), \bibinfo{edition}{2nd} ed.

\bibitem[{\citenamefont{Festa and
  Galleani~d'Agliano}(1978)}]{festa-dagliano-1978}
\bibinfo{author}{\bibfnamefont{R.}~\bibnamefont{Festa}} \bibnamefont{and}
  \bibinfo{author}{\bibfnamefont{E.}~\bibnamefont{Galleani~d'Agliano}},
  \bibinfo{journal}{Physica A} \textbf{\bibinfo{volume}{90A}},
  \bibinfo{pages}{229} (\bibinfo{year}{1978}).

\bibitem[{\citenamefont{Alexander et~al.}(1981)\citenamefont{Alexander,
  Bernasconi, Schneider, and Orbach}}]{Alexander-Orbach-1981}
\bibinfo{author}{\bibfnamefont{S.}~\bibnamefont{Alexander}},
  \bibinfo{author}{\bibfnamefont{J.}~\bibnamefont{Bernasconi}},
  \bibinfo{author}{\bibfnamefont{W.~R.} \bibnamefont{Schneider}},
  \bibnamefont{and} \bibinfo{author}{\bibfnamefont{R.}~\bibnamefont{Orbach}},
  \bibinfo{journal}{Rev. Mod. Phys.} \textbf{\bibinfo{volume}{53}},
  \bibinfo{pages}{175} (\bibinfo{year}{1981}),
  \urlprefix\url{http://link.aps.org/abstract/RMP/v53/p175}.

\bibitem[{\citenamefont{Golden et~al.}(1985)\citenamefont{Golden, Goldstein,
  and Lebowitz}}]{golden-1985}
\bibinfo{author}{\bibfnamefont{K.}~\bibnamefont{Golden}},
  \bibinfo{author}{\bibfnamefont{S.}~\bibnamefont{Goldstein}},
  \bibnamefont{and} \bibinfo{author}{\bibfnamefont{J.~L.}
  \bibnamefont{Lebowitz}}, \bibinfo{journal}{Phys. Rev. Lett.}
  \textbf{\bibinfo{volume}{55}}, \bibinfo{pages}{2629} (\bibinfo{year}{1985}),
  \urlprefix\url{http://link.aps.org/abstract/PRL/v55/p2629}.

\bibitem[{\citenamefont{Schneider et~al.}(1988)\citenamefont{Schneider, Politi,
  and S{\"o}rensen}}]{schneider-1988}
\bibinfo{author}{\bibfnamefont{T.}~\bibnamefont{Schneider}},
  \bibinfo{author}{\bibfnamefont{A.}~\bibnamefont{Politi}}, \bibnamefont{and}
  \bibinfo{author}{\bibfnamefont{M.~P.} \bibnamefont{S{\"o}rensen}},
  \bibinfo{journal}{Phys. Rev. A} \textbf{\bibinfo{volume}{37}},
  \bibinfo{pages}{948} (\bibinfo{year}{1988}),
  \urlprefix\url{http://link.aps.org/abstract/PRA/v37/p948}.

\bibitem[{\citenamefont{Larkin and
  Ovchinnikov}(1973)}]{larkin-ovchinnikov-1974}
\bibinfo{author}{\bibfnamefont{A.~I.} \bibnamefont{Larkin}} \bibnamefont{and}
  \bibinfo{author}{\bibfnamefont{Y.~N.} \bibnamefont{Ovchinnikov}},
  \bibinfo{journal}{Zh. Eksp. Teor. Fiz.} \textbf{\bibinfo{volume}{65}},
  \bibinfo{pages}{704} (\bibinfo{year}{1973}), \bibinfo{note}{[Sov. Phys. JETP
  38, 854 (1974)]}.

\bibitem[{\citenamefont{Vinokur et~al.}(1990)\citenamefont{Vinokur, Feigel'man,
  Geshkenbein, and Larkin}}]{vinokur-vortex-liquid-1990}
\bibinfo{author}{\bibfnamefont{V.~M.} \bibnamefont{Vinokur}},
  \bibinfo{author}{\bibfnamefont{M.~V.} \bibnamefont{Feigel'man}},
  \bibinfo{author}{\bibfnamefont{V.~B.} \bibnamefont{Geshkenbein}},
  \bibnamefont{and} \bibinfo{author}{\bibfnamefont{A.~I.}
  \bibnamefont{Larkin}}, \bibinfo{journal}{Phys. Rev. Lett.}
  \textbf{\bibinfo{volume}{65}}, \bibinfo{pages}{259} (\bibinfo{year}{1990}),
  \urlprefix\url{http://link.aps.org/abstract/PRL/v65/p259}.

\bibitem[{\citenamefont{Smelyanskiy et~al.}(1997)\citenamefont{Smelyanskiy,
  Dykman, Rabitz, and Vugmeister}}]{smelyanskiy-dykman-1997}
\bibinfo{author}{\bibfnamefont{V.~N.} \bibnamefont{Smelyanskiy}},
  \bibinfo{author}{\bibfnamefont{M.~I.} \bibnamefont{Dykman}},
  \bibinfo{author}{\bibfnamefont{H.}~\bibnamefont{Rabitz}}, \bibnamefont{and}
  \bibinfo{author}{\bibfnamefont{B.~E.} \bibnamefont{Vugmeister}},
  \bibinfo{journal}{Phys. Rev. Lett.} \textbf{\bibinfo{volume}{79}},
  \bibinfo{pages}{3113} (\bibinfo{year}{1997}),
  \urlprefix\url{http://link.aps.org/abstract/PRL/v79/p3113}.

\bibitem[{\citenamefont{Stepanow and Sommer}(1990)}]{stepanov-sommer-1990}
\bibinfo{author}{\bibfnamefont{S.}~\bibnamefont{Stepanow}} \bibnamefont{and}
  \bibinfo{author}{\bibfnamefont{J.-U.} \bibnamefont{Sommer}},
  \bibinfo{journal}{J. Phys. A: Math. Gen.} \textbf{\bibinfo{volume}{23}},
  \bibinfo{pages}{L541} (\bibinfo{year}{1990}),
  \urlprefix\url{http://stacks.iop.org/0305-4470/23/L541}.

\bibitem[{\citenamefont{Stepanow}(1991)}]{stepanov-1991}
\bibinfo{author}{\bibfnamefont{S.}~\bibnamefont{Stepanow}},
  \bibinfo{journal}{Phys. Rev. A} \textbf{\bibinfo{volume}{43}},
  \bibinfo{pages}{2771} (\bibinfo{year}{1991}),
  \urlprefix\url{http://link.aps.org/abstract/PRA/v43/p2771}.

\end{thebibliography}

\end{document}